\documentclass{article}
\usepackage{spconf,amsmath,graphicx}
\usepackage{amsfonts,makecell,multirow,enumitem,booktabs,cite,tabularx}

\title{LEARNING ACOUSTIC WORD EMBEDDINGS WITH PHONETICALLY ASSOCIATED TRIPLET NETWORK}
\name{Hyungjun Lim, Younggwan Kim, Youngmoon Jung, Myunghun Jung, and Hoirin Kim}
\address{School of Electrical Engineering, Korea Advanced Institute of Science and Technology, \\
	 Daejeon, Republic of Korea\\
	\normalsize\texttt{\{hyungjun.lim, cleanthink, dudans, kss2517, hoirkim\}@kaist.ac.kr}}

\begin{document}
%
\maketitle
\begin{abstract}
Previous researches on acoustic word embeddings used in query-by-example spoken term detection have shown remarkable performance improvements when using a triplet network. However, the triplet network is trained using only a limited information about acoustic similarity between words. In this paper, we propose a novel architecture, \textit{phonetically associated triplet network} (PATN), which aims at increasing discriminative power of acoustic word embeddings by utilizing phonetic information as well as word identity. The proposed model is learned to minimize a combined loss function that was made by introducing a cross entropy loss to the lower layer of LSTM-based triplet network. We observed that the proposed method performs significantly better than the baseline triplet network on a word discrimination task with the WSJ dataset resulting in over $20\%$ relative improvement in recall rate at $1.0$ false alarm per hour. Finally, we examined the generalization ability by conducting the out-of-domain test on the RM dataset.  

\end{abstract}

%

\section{Introduction}
\label{sec:intro}

Keyword spotting (KWS), sometimes also referred to as spoken term detection (STD), is one of the most widely used speech-related technique which aims at detecting the occurrence of a particular word or multi-word phrases in a given speech utterance. A prominent example of STD is wake-up word detection (WWD), embedded in ``Google's voice search'' \cite{schalkwyk2010}, ``Apple's Siri'', and ``Amazon's Echo'', which has much attention recently. 

A conventional approach for STD is based on the keyword/filler hidden Markov model (HMM) \cite{rose1990, wilpon1990, wilpon1991}, which remains strongly competitive until these days \cite{panchapagesan2016}. At runtime, Viterbi decoding is used to search the best path in the decoding graph, which can be computationally expensive depending on the HMM topology. 
Other approaches to STD rely on pattern matching schemes such as dynamic time warping (DTW) \cite{sakoe1978} to quantify the similarity between templates, including Gaussian posteriorgrams \cite{zhang2009}, phoneme posteriograms \cite{hazen2009}, and CNN-based bottleneck features \cite{lim2017}. However, DTW has known inadequacies \cite{rabiner1978} and is quadratic-time in the duration of the segments \cite{levin2013}.

Recently, many researchers have tried to embed the variable length of speech signal into a fixed dimensional vector called \textit{acoustic word embeddings} \cite{chen2015, hou2016, kamper2016, wang2018, Shan2018, settle2016, settle2017} which can easily measure the similarity between them by using cosine distance with only a small amount of computations compare to the DTW. In \cite{chen2015, hou2016}, acoustic word embeddings were generated from a long short-term memory (LSTM) trained with whole word output targets, obtaining better performance than other DTW-based approaches. Later, in \cite{settle2016,settle2017}, an additional performance gain was achieved by introducing a \textit{triplet network} \cite{hoffer2015} which was originally proposed in \cite{bromley1994} to solve the problem of signature verification. The network was optimized by using a triplet loss function on the LSTM output layer that tried to maximize the distance between embeddings from the same word classes and simultaneously minimize the distance between embeddings from the different word classes. However, the triplet loss did not consider phonetic information but rely only on relative relationship between words.

In this paper, we propose a \textit{phonetically associated triplet network} (PATN) that expand the previous work through a hierarchical multitask learning scheme \cite{toshniwal2017, krishna2018} to utilize phonetic information in the triplet network. Similar to \cite{toshniwal2017, krishna2018}, a frame-level cross entropy loss function is introduced to the lower layer of the triplet network to explicitly impose the concept that different layers encode different levels of information; the lower layer models the frame-level variations while the higher layer describes the relationship among words. Experimental results show that, more discriminative embeddings can be obtained from the proposed model trained on a convex combination of the two loss functions.

The remainder of the paper is organized as follows. In Section \ref{sec:model} we describe the proposed modeling strategy in detail followed by experimental results and analysis in Section \ref{sec:exp}. Finally, our conclusions and future directions are summarized in Section \ref{sec:conclusion}.

\section{MODEL}
\label{sec:model}
Denote an input acoustic feature sequence as $ \mathbf{x} = ( \mathbf{x}_1, ... \, , \mathbf{x}_T )$ and corresponding labels as $y=(y_1, ... \, , y_T)$ where $T$ is the number of acoustic frames in a word. In this section, we first review previous work (Figure \ref{fig:model}-(a)) and then present our proposed method in detail (Figure \ref{fig:model}-(b)). 

\subsection{Triplet network}
\label{ssec:triplet}
A Triplet network \cite{hoffer2015} consists of three identical networks with shared weights. Intuitively, the triplet network encourages to find an embedding space where the distances between examples from the same word class ($i.e.$, $\mathbf{x}^{A(\text{anchor})}$ and $\mathbf{x}^{S(\text{same})}$) are smaller than those from different word classes ($i.e.$, $\mathbf{x}^{A(\text{anchor})}$ and $\mathbf{x}^{D(\text{different})}$) by at least a margin $m$. Formally, given a triplet of acoustic feature sequences $\mathbf{X}=\{\mathbf{x}^A, \, \mathbf{x}^S, \, \mathbf{x}^D\}$, the triplet network is trained to minimize triplet loss defined as
\begin{align}
	\mathcal{L}_{T} = \max {\{ 0, \: m + d_+ - d_- \}} \\
	d_+ = 1 - \cos(f(\mathbf{x}^A), \, f(\mathbf{x}^S)) \\
	d_- = 1 - \cos(f(\mathbf{x}^A), \, f(\mathbf{x}^D)) 
	\label{eq1}
\end{align}
\noindent where $f(\cdot)$ is an acoustic word embedding function, $m$ is the margin constraint, and $d_+$ and $d_-$ are the cosine distance between acoustic word embeddings belong to same/different word classes, respectively. As in \cite{settle2016, settle2017}, we use the concatenation of the hidden representations from a bidirectional LSTM network \cite{schuster1997} as our acoustic embedding function. 

\subsection{Phonetically associated triplet network (PATN)}
\label{ssec:TASN}
\begin{figure}[tb]
	\centering
	\centerline{\includegraphics[width=8.5cm]{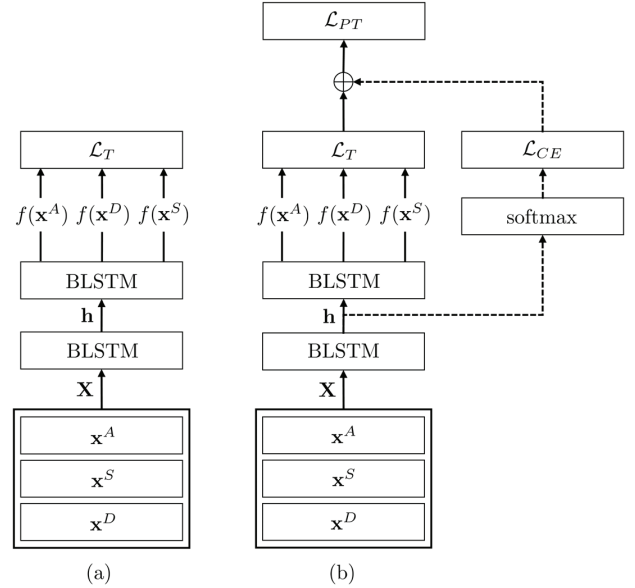}}
	\vspace{0.1cm}
	\caption{Block diagrams of the (a) triplet network and (b) phonetically associated triplet network.}
	\vspace{0.3cm}
	\label{fig:model}
\end{figure}

Since the triplet network trained on word-level criterion, the resulting acoustic word embedding may not be sensitive to small amount of variation within words. Thus we propose the phonetically associated triplet network (PATN) which is jointly trained on both word- and frame-level criteria and given by
\begin{equation}
	\mathcal{L}_{PT} = (1-\lambda)\mathcal{L}_{T} + \lambda \mathcal{L}_{CE} 
	\label{eq2}
\end{equation}
\begin{equation}
\mathcal{L}_{CE} = - \sum_{n=1}^N \sum_{i=1}^{C} I\{y_n = i\}\log{ 
	\frac{ e^{ \mathbf{W}^{T}_{i} \mathbf{h}_{n} + b_i }}
	{ \sum_{j=1}^{C} e^{ \mathbf{W}^{T}_{j} \mathbf{h}_{n} + b_j }}}
\label{eq3}
\end{equation}
\noindent where $\mathcal{L}_{CE}$ is a cross-entropy loss obtained from the softmax with $C$ classes, $\lambda$ is a hyper-parameter which controls the trade-off between the two loss functions, $N$ is the number of data in a mini-batch, and the indicator function $I(y_n = i)$ is $1$ if $i$ is equal to class label $y_n$ and $0$ for otherwise. Similar to \cite{toshniwal2017, krishna2018}, we introduce the cross-entropy loss function to the lower layer of the triplet network as depicted in Figure \ref{fig:model}-(b). Such a low-level auxiliary task explicitly encourages intuitive and empirical observation that different layers encode different levels of information.

\section{EXPERIMENTS}
\label{sec:exp}
To confirm whether the embeddings extracted from the proposed architecture represent the characteristics of words, we conducted the experiments of word discrimination \cite{settle2016, settle2017, carlin2011, kamper2016}, which is a simplified version of wake-up word detection where the word boundary information is given. Similar to wake-up word detection, the test of word discrimination consists of two steps: \textit{enrollment} and \textit{verification}. In the enrollment phase, a speech segment of query is fed into the trained BLSTM. Then, the enrollment embedding is generated by concatenating the two last hidden state vectors from forward and backward directions of BLSTM. In the verification phase, test embeddings are generated in the same way followed by measuring the cosine distance between the enrollment and test embeddings. To get more reliable results, we used averaged cosine distance calculated from 5 enrollment queries for each keyword. By sweeping a threshold, we can obtain the recall at a certain point of false alarm which is usually used to measure the performance in wake-up word detection task \cite{chen2014, chen2015, sainath2015}.

\subsection{Datasets}
\label{ssec:data}

Our models were trained on $100k$ triplets selected from WSJ \cite{paul1992} SI-284 training set. The triplets consisted of three word segments which were randomly chosen from entire words in training set. Note that the minimum duration of the segments was $0.5$ sec and the segments were extracted by using the forced alignment of the transcriptions from the GMM-HMM acoustic model trained on the same training set using the open-source Kaldi toolkit \cite{povey2011}. For testing, we used two kinds of datasets: test sets (\textit{i.e.}, eval92 and eval93) from WSJ database (in-domain test) and training set from RM database \cite{price1988} (out-of-domain test). We selected queries from the each test set with high frequency of occurrence which are listed in Table \ref{tab:list}. 

\subsection{Model details}
\label{ssec:model}
We represented the speech signal using $40$-dimensional Mel-filterbank log energy which was calculated from $25$ msec frame size with $50\%$ overlap. The state-level label on each speech frame was generated by the GMM-HMM acoustic model through forced alignment for the PATN training. Note that we utilized both monophone ($132$ states) and tied-triphone ($3342$ states) state-level labels. The acoustic word embeddings were generated by concatenating two last hidden state vectors from forward and backward directions of BLSTMs which consisted with $2$ hidden layers and $128$ hidden units. All the models were trained for $40$ epochs using the Adam optimization algorithm \cite{kingma2015} which was implemented in tensorflow toolkit \cite{tensorflow2015-whitepaper} with a batch size of $128$, learning rate of $0.0005$, $\beta_1=0.9$, $\beta_2=0.999$, and $\epsilon=1\cdot10^{-8}$. 

\begin{figure}[tb]
	\centering
	\centerline{\includegraphics[width=8.5cm]{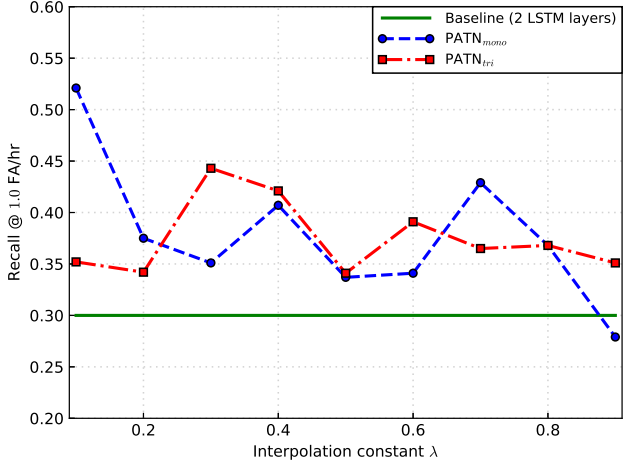}}
	\caption{Effect of varying the interpolation constant $\lambda$ in the WSJ development set (\textit{i.e.}, dev93). $\text{PATN}_{mono}$ and $\text{PATN}_{tri}$ indicate that ``monophone'' and ``tied-triphone'' state-level labels are used in  cross-entropy loss function, respectively. }
	\vspace{0.3cm}
	\label{fig:graph}
\end{figure}

\subsection{Results}
\label{ssec:results}

We first examined how the mixing weight between the two terms in the PATN loss affects performance. For this, we measured word discrimination performance in the development data in terms of recall at the operating threshold of $1.0$ false alarm (FA) per hour. As can be seen in Figure \ref{fig:graph}, the proposed method with any value of $\lambda \neq1$ outperformed the baseline, with the best performance obtained at $\lambda=0.1$ in $\text{PATN}_{mono}$. This means that the additional phonetic information, especially when we use the monophone state-level targets, can improve the performance of the triplet network.   

Next, we summarize the performance of baseline and proposed method measured in both in-domain and out-of-domain test sets which is depicted in Table \ref{tab:exp1}. Here, we used the $\lambda$ that achieved the highest performance in the development set (see Figure \ref{fig:graph}). We can clearly see that our proposed method outperformed the baseline as in the previously observed results from the development set even though we did not increase the model size. Surprisingly, our proposed method was still effective with out-of-domain environment, achieving over $20\%$ relative improvement with the same model size.

\begin{table}[tb]
	\caption{List of queries used for evaluation. ID and OOD represent that in-domain and out-of-domain, respectively.}
		\vspace{0.4cm}
	\label{tab:list}
	\centering
	\begin{tabularx}{\linewidth}{ >{\hsize=0.15\hsize}c >{\hsize=0.85\hsize}X }
		\Xhline{3\arrayrulewidth}
		WSJ (ID) & company, dollars, from, hundred, nineteen, percent, point, seven \\
		\hline
		RM (OOD) & Bismark, coral, displacement, Formosa, frigates, kilometers, Mozambique, Siberian, Thailand, Tonkin, Westpac, Zulu \\
		\Xhline{3\arrayrulewidth}
	\end{tabularx}
\end{table}

\begin{table}[tb]
	\caption{Performance summarization of the baseline triplet network and phonetically associated triplet network on in-domain and out-of-domain test sets.}
	\vspace{0.4cm}
	\label{tab:exp1}
	\centering
	\begin{tabularx}{\linewidth}{ >{\hsize=0.52\hsize}X >{\hsize=0.24\hsize}c >{\hsize=0.24\hsize}c }
		\Xhline{3\arrayrulewidth}
		\hfil \multirow{2}*{\textbf{Model}} \hfill & \multicolumn{2}{c}{\textbf{Recall @ $1.0$ FA/hr}} \\
		\cline{2-3} & WSJ (ID) & RM (OOD)\\		
		\Xhline{3\arrayrulewidth}
		\textbf{Baseline} \cite{settle2016} & & \\
		\qquad TN ($1$ BLSTM layer)  &   $0.436$ &  $0.463$ \\
		\qquad TN ($2$ BLSTM layers) &   $0.554$ &  $0.485$ \\
		\qquad TN ($3$ BLSTM layers) &   $0.559$ &  $0.548$ \\				
		\hline
		\textbf{Proposed} ($2$ BLSTM layers) & & \\
		\qquad $\text{PATN}_{mono}$ ($\lambda=0.1$)  		&   $\mathbf{0.714}$ 	&  $\mathbf{0.582}$ \\
		\qquad $\text{PATN}_{tri}$ \quad ($\lambda=0.3$)   	&  $0.660$ 				&  $0.507$ \\
		
		\Xhline{3\arrayrulewidth}
	\end{tabularx}
\vspace{0.3cm}
\end{table}

\subsection{Analysis}
\label{ssec:analysis}
\begin{figure*}[htb]
	
	\begin{minipage}[t]{.5\linewidth}
		\centering
		\centerline{\includegraphics[width=\linewidth]{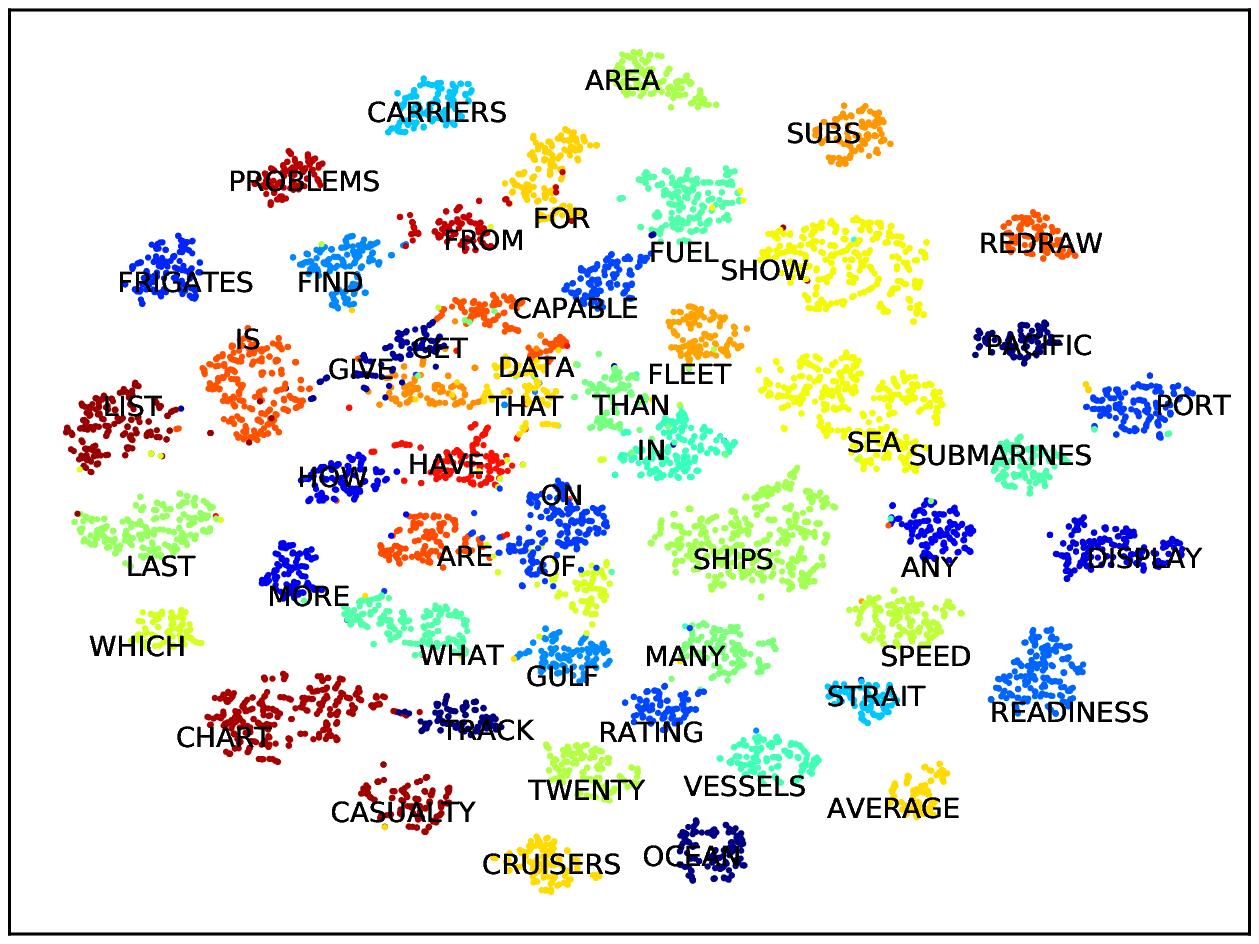}}
				\vspace{0.3cm}
		\centerline{(a) Triplet network}\medskip
	\end{minipage}
	\begin{minipage}[t]{.5\linewidth}
		\centering
		\centerline{\includegraphics[width=\linewidth]{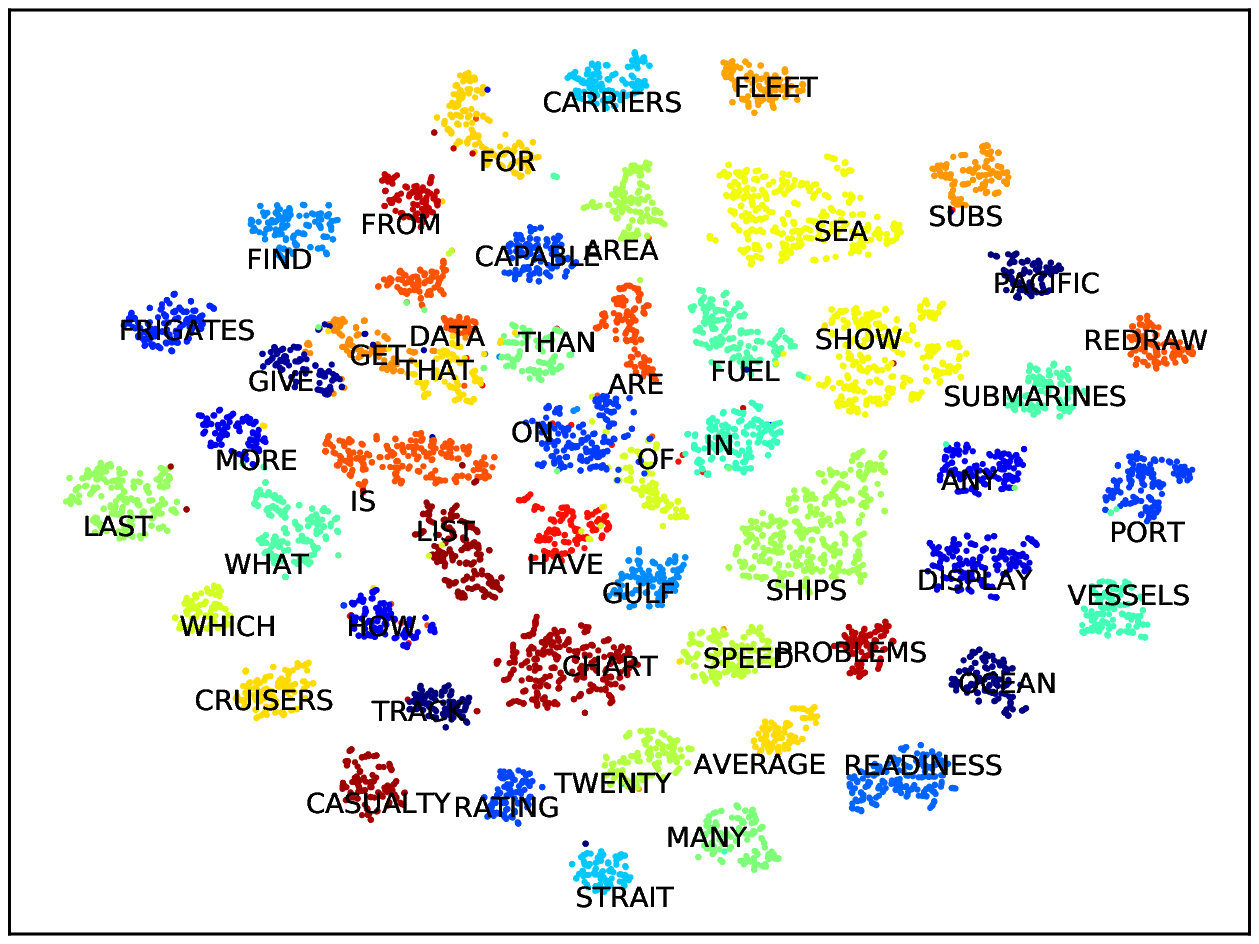}}
				\vspace{0.3cm}
		\centerline{(b) Phonetically associated triplet network}\medskip
	\end{minipage}
	\caption{t-SNE\cite{maaten2008} visualization of acoustic word embeddings of the out-of-domain test set produced by the (a) triplet network and (b) phonetically associated triplet network (Best viewed in color). }
	\vspace{0.3cm}
	\label{fig:analysis}
\end{figure*} 
To verify the effectiveness of the proposed method, we plotted a two-dimensional visualization of embeddings extracted from words appeared at least $100$ occurrences of frequency in the out-of-domain test set via t-distributed stochastic neighbor embedding (t-SNE) \cite{maaten2008}, which is a non-linear dimensionality reduction technique particularly well suited for high-dimensional data (see Figure \ref{fig:analysis}). As you can see, most of the embeddings are clustered into their corresponding word classes in both the baseline and the proposed method. Since the triplet network was learned based on the similarity between words, we can obtain the discriminative word representations in the embedding space even unused data in the training. We can also observe that a confusability between words was relaxed in our proposed method resulting more separable clusters of word embeddings (\textit{e.g.}, \texttt{give} vs. \texttt{get}, \texttt{have} vs. \texttt{how}, \texttt{for} vs. \texttt{from}, \texttt{chart} vs. \texttt{track}, etc.). Therefore, we can conclude that the proposed method can increase the discrimination between words while maintaining the generalization power of the triplet network.

\section{CONCLUSION}
\label{sec:conclusion}
In this paper, we proposed a novel architecture called \textit{phonetically associated triplet network} (PATN) which can learn more discriminative embeddings by inserting phonetic information into the triplet network. In the method, we applied the hierarchical multitask learning framework to the triplet network by introducing an auxiliary cross-entropy loss function at the lower layer of the LSTM. On the same-different word discrimination task, which is similar to wake-up word detection except word boundary is given, our approach outperformed the previous triplet network architecture, achieving over $20\%$ relative improvement in terms of recall at the operating threshold of $1.0$ false alarm (FA) per hour. Moreover, we showed that our model could generalize their performance in the out-of-domain dataset. Finally, we have demonstrated that the phonetic information is really helpful to generate acoustic word embeddings through qualitative comparison of proposed method and the baseline with t-SNE visualizations. 

As a future direction, we will expand our works by using large amount of training data to improve performance of the triphone based PATN which was mentioned in Section \ref{ssec:results}. To do so, we also look into ways of improving the extremely long training times such as triplet selection \cite{schroff2015} and class-wise triplet loss \cite{ming2017}. Based on not only the promising results from the out-of-domain task but also further considerations like as temporal context information \cite{Yuan2018}, our method may successfully be applied to the personalized wake-up word detection task.


\section{ACKNOWLEDGEMENT}
\label{sec:ack}

This material is based upon work supported by the Ministry of Trade, Industry \& Energy (MOTIE, Korea) under Industrial Technology Innovation Program (No.10063424, Development of distant speech recognition and multi-task dialog processing technologies for in-door conversational robots).

\bibliographystyle{IEEEbib}
{\footnotesize
\bibliography{refs}}

\end{document}